\documentclass[aps,prb,reprint,amsmath,amssymb,nosuperscriptaddress]{revtex4-1}
\usepackage{bm}
\usepackage[colorlinks]{hyperref}
\usepackage{mathrsfs}

\newcommand\de{\partial}
\newcommand\dd{\mathrm{d}}
\newcommand\La{\mathscr{L}}
\newcommand\PD[2]{\frac{\de #1}{\de #2}}
\newcommand\vek[1]{\bm{#1}}
\newcommand\imag{\mathrm{i}}
\let\eps=\epsilon

\begin{document}
\title{Noether currents of locally equivalent symmetries}

\author{Tom\'{a}\v{s} Brauner}
\email{tomas.brauner@uis.no}
\affiliation{Department of Mathematics and Physics, University of Stavanger, 4036 Stavanger, Norway}

\begin{abstract}
Local symmetry transformations play an important role for establishing the existence and form of a conserved (Noether) current in systems with a global continuous symmetry. We explain how this fact leads to the existence of linear relations between Noether currents of distinct global symmetries that coincide on the local level, thus generalizing the well-known relationship $\vek L=\vek r\times\vek p$ between momentum $\vek p$ and angular momentum $\vek L$. As a byproduct, we find a natural interpretation for the discrepancy between the canonical and metric energy-momentum tensors in theories of particles with spin. A symmetric energy-momentum tensor can thus be obtained from the Noether procedure without adding any \emph{ad hoc} corrections or imposing additional constraints such as gauge invariance in Maxwell's electrodynamics.
\end{abstract}

\maketitle


\section{Introduction}
\label{sec:intro}

The link between symmetry and conservation laws, discovered by Noether,\cite{Noether:1918zz} underpins much of our understanding of the fundamental laws of nature. An excellent historical overview of the subject was given by Ref.~\onlinecite{Kosmann:2011}, to which we refer the reader for more details; see also Ref.~\onlinecite{Neuenschwander:2011} for a gentle introduction.

In this note, we revisit the application of Noether's ideas to classical field theory. Our starting point is a simplified derivation of the Noether current associated with a given global symmetry of the action, in which one identifies the current with the help of a local, coordinate-dependent transformation. This approach, sometimes attributed to Gell-Mann and L\'evy,\cite{GellMann:1960np} is well established in standard textbooks on quantum field theory.\cite{Itzykson:1980,*Weinberg:1995v1}

We consider the general class of theories whose action $S$ is expressed as a spacetime integral of a Lagrangian density $\La$ that itself is a local function of a set of fields $\phi_A$ and their derivatives. Suppose that the action is invariant under some \emph{global} transformation of both the fields and spacetime coordinates, denoted collectively as $x$. Now perform an infinitesimal transformation whose parameter $\eps(x)$ is allowed to depend on the coordinates, only assuming that for constant $\eps(x)=\eps$ the transformation reduces to the assumed global symmetry. The variation of the action under such a transformation then necessarily depends on $\eps(x)$ only through its derivatives, and by integration by parts can be brought to the form\footnote{For the sake of simplicity, we will ignore boundary terms arising from integration by parts. This can be justified for instance by assuming that the integration is performed over the whole spacetime and all the fields drop to zero sufficiently fast at infinity.}
\begin{equation}
\delta S=\int\dd x\,J^\mu(x)\de_\mu\eps(x).
\label{master}
\end{equation}
Here $J^\mu(x)$ is the Noether current associated with the assumed global symmetry. The Hamilton principle dictates that for fields satisfying the equation of motion (``on-shell'' fields), $\delta S=0$ for \emph{any} infinitesimal variation of the fields, in particular for the one induced by the performed local transformation. This implies that the Noether current is conserved on-shell, $\de_\mu J^\mu=0$.\footnote{We use relativistic index notation along with Einstein's summation convention, but we make no \emph{a priori} assumptions about the spacetime symmetry: all the general results presented below are valid for relativistic and nonrelativistic systems alike.}

The master equation~\eqref{master} constitutes the starting point for the rest of the paper. In Sec.~\ref{sec:explicit} we discuss ambiguities in the definition of the Noether current, and list several explicit expressions for the current under increasingly relaxed assumptions. This section does not contain any new material, yet it offers a more general treatment than many introductory texts on the subject. The core of the paper consists of Secs.~\ref{sec:relations} and~\ref{sec:gauge}, which approach the relations between Noether currents of locally identical symmetries\cite{Brauner:2014aha} following two similar but complementary methodologies. A collection of examples is worked out in Sec.~\ref{sec:examples}. The important case of the energy-momentum (EM) tensor is covered separately in Sec.~\ref{sec:EMtensor}.


\section{Some explicit expressions}
\label{sec:explicit}

It is clear from Eq.~\eqref{master} that the Noether current is only defined up to addition of terms whose divergence vanishes for all field configurations (``off-shell''). Moreover, adding a term to $J^\mu$ that vanishes identically on-shell has no effect on the ensuing conservation law. The same conservation law therefore corresponds to a whole equivalence class of currents, $J^\mu+\bar J_1^\mu+\bar J_2^\mu$, where $\de_\mu\bar J_1^\mu=0$ off-shell and $\bar J_2^\mu=0$ on-shell.\cite{Gordon:1984xb} We will demonstrate below that the ambiguity of the Noether current with respect to $\bar J^\mu_2$ can be traced to the ambiguity in the choice of the local transformation, employed to produce the variation~\eqref{master}. This requires, however, a sufficiently general notion of a local symmetry transformation. The ambiguity with respect to $\bar J^\mu_1$, on the other hand, is inevitable when Eq.~\eqref{master} is used to define the Noether current.

To start, let us at first for the sake of simplicity assume that the Lagrangian density $\La$ depends only on the fields $\phi_A$ and their first derivatives $\de_\mu\phi_A$, $\La=\La(\phi,\de_\mu\phi,x)$, where the argument $x$ indicates possible explicit coordinate dependence. Now perform the following infinitesimal local transformation of the fields and coordinates,
\begin{equation}
\begin{split}
\phi_A'(x')&=\phi_A(x)+\eps(x)\xi_A(x),\\
x'^\mu&=x^\mu+\eps(x)\omega^\mu(x).
\end{split}
\label{transfo}
\end{equation}
Under this transformation, the action varies by\footnote{In $\de_\mu\La$, the partial derivative is understood to act just on the explicit coordinate dependence of the Lagrangian.}
\begin{equation}
\begin{split}
\delta S={}&\int\dd x\,\biggl\{\La\de_\mu(\eps\omega^\mu)+\eps\omega^\mu\de_\mu\La+\PD\La{\phi_A}\eps\xi_A\\
&+\PD\La{(\de_\mu\phi_A)}\bigl[\de_\mu(\eps\xi_A)-\de_\mu(\eps\omega^\nu)\de_\nu\phi_A\bigr]\biggr\}.
\end{split}
\end{equation}
The first term comes from the Jacobian of the coordinate transformation, the second from the explicit coordinate dependence of the Lagrangian, the third from variation of $\phi_A$, and the last from variation of $\de_\mu\phi_A$. Upon collecting terms proportional to $\eps(x)$ and $\de_\mu\eps(x)$, we readily extract both the condition for the existence of global symmetry of the action and the corresponding Noether current. The former reads
\begin{multline}
\PD\La{\phi_A}\xi_A+\PD\La{(\de_\mu\phi_A)}\de_\mu\xi_A+\de_\mu(\La\omega^\mu)\\
-\PD\La{(\de_\mu\phi_A)}(\de_\nu\phi_A)\de_\mu\omega^\nu=\de_\mu K^\mu,
\label{invariance}
\end{multline}
where $K^\mu(x)$ is some vector function. In the following, we will refer to Eq.~\eqref{invariance} as the ``invariance condition.'' The Noether current then takes the form
\begin{equation}
J^\mu=\PD\La{(\de_\mu\phi_A)}\xi_A+\left[\delta^\mu_\nu\La-\PD\La{(\de_\mu\phi_A)}\de_\nu\phi_A\right]\omega^\nu-K^\mu.
\label{noether}
\end{equation}
Note that despite the limitation on the number of derivatives acting on the fields in the Lagrangian, this result is already pretty general. First, it allows for explicit coordinate dependence of the Lagrangian. Second, we did \emph{not} require that the functions $\xi_A(x)$ depend only on the fields and $\omega^\mu(x)$ only on the coordinates. Both of these can depend on the coordinates, on the fields as well as on their derivatives.

It is now easy to check that both the invariance condition~\eqref{invariance} and the Noether current~\eqref{noether} remain unchanged under the following simultaneous replacements
\begin{equation}
\begin{gathered}
\omega^\mu\to\omega^\mu+\bar\omega^\mu,\qquad
\xi_A\to\xi_A+\bar\omega^\mu\de_\mu\phi_A,\\
K^\mu\to K^\mu+\La\bar\omega^\mu,
\end{gathered}
\label{gauge_freedom}
\end{equation}
where $\bar\omega^\mu$ is an arbitrary function of the coordinates, fields and their derivatives. This indicates a redundancy in the description of the \emph{same} physical symmetry: formally a whole family of transformations gives the same Noether current.\cite{Candotti:1970dc} The reason for this is that the spacetime coordinate $x^\mu$ is merely a dummy variable that is integrated over. The action should rather be thought of as a functional of the fields only; redefinitions of the coordinate cannot have any physical content.\cite{Banados:2016zim} In particular, one can always eliminate $\omega^\mu$ by setting $\bar\omega^\mu=-\omega^\mu$, and thereby arrive at an equivalent description of the same physical symmetry in terms of a transformation of $\phi_A$ alone,
\begin{equation}
\phi_A'(x)=\phi_A(x)+\eps(x)\xi_A(x)-\eps(x)\omega^\mu(x)\de_\mu\phi_A(x).
\label{transfo2}
\end{equation}
This underlines the importance of symmetry transformations that depend on the derivatives of the fields.

We will now generalize the discussion by allowing for Lagrangians containing arbitrarily high derivatives of the fields. To simplify the result, we will set $\omega^\mu=0$, as this can always be absorbed into a redefinition of $\xi_A$ as we just argued. The invariance condition on the Lagrangian then takes a very simple form, generalizing Eq.~\eqref{invariance},
\begin{equation}
\sum_{n=0}^\infty\frac{\de\La}{\de(\de_{\mu_1}\dotsb\de_{\mu_n}\phi_A)}\de_{\mu_1}\dotsb\de_{\mu_n}\xi_A=\de_\mu K^\mu.
\label{invariance_general}
\end{equation}
By inspecting the variation of the action under a local symmetry transformation, we then get the corresponding expression for the Noether current. After some manipulation, one thus finds
\begin{align}
\label{noether_general}
&J^{\mu}=-K^{\mu}+\sum_{n=1}^\infty\sum_{k=1}^n(-1)^{k+1}\\
\notag
&\times\left[\de_{\mu_2}\dotsb\de_{\mu_k}\frac{\de\La}{\de(\de_{\mu}\de_{\mu_2}\dotsb\de_{\mu_n}\phi_A)}\right](\de_{\mu_{k+1}}\dotsb\de_{\mu_n}\xi_A).
\end{align}
This result dates back at least to 1966,\cite{Candotti:1970dc,Dass:1966zza,*Rosen:1972ku} even though higher-order derivatives of fields were implicitly allowed already in the original work of Noether.\cite{Noether:1918zz}

For yet another line of generalization, note that a general local symmetry transformation can be cast as a series in the derivatives of the parameter $\eps(x)$. The leading term with no derivatives, displayed in Eqs.~\eqref{transfo} or~\eqref{transfo2}, defines the corresponding global symmetry. Higher-order contributions, which by construction do not affect the global symmetry, may however be present. To see what this implies for the Noether current, let us replace the first line of Eq.~\eqref{transfo} with
\begin{equation}
\begin{split}
\phi_A'(x')={}&\phi_A(x)+\eps(x)\xi_A(x)\\
&+\sum_{n=1}^\infty\sigma_A^{\mu_1\dotsb\mu_n}(x)\de_{\mu_1}\dotsb\de_{\mu_n}\eps(x),
\end{split}
\label{transfosigma}
\end{equation}
where $\sigma_A^{\mu_1\dotsb\mu_n}(x)$ is a set of tensor functions of the coordinates, fields and their derivatives. Since the added terms only depend on the derivatives of $\eps(x)$, they do not affect the invariance condition on the action. Upon some integration by parts, one can directly extract the corresponding contribution to the Noether current,
\begin{equation}
J^{\mu}= J^{\mu}\Bigr|_{\sigma=0}+\sum_{n=1}^\infty(-1)^{n+1}\de_{\mu_2}\dotsb\de_{\mu_n}\left(\sigma_A^{\mu\mu_2\dotsb\mu_n}\frac{\delta S}{\delta\phi_A}\right),
\label{Jsigma}
\end{equation}
where $\delta S/\delta\phi_A$ is the variation of the action with respect to $\phi_A$. As the latter defines the equation of motion, it is obvious that the new contributions to the current vanish on-shell for any choice of $\sigma_A^{\mu_1\dotsb\mu_n}$.

This finishes our discussion of ambiguities in Noether currents and the associated local symmetry transformations. Equations~\eqref{noether}, \eqref{noether_general} and~\eqref{Jsigma} will serve as a reference in Secs.~\ref{sec:examples} and~\ref{sec:EMtensor} where we work out concrete examples.


\section{Relations among currents}
\label{sec:relations}

Some (typically spacetime) symmetries, albeit distinct globally, may not be distinguishable locally.\cite{Low:2001bw} The simplest example is that of spatial rotations and translations, which locally both correspond to a coordinate shift. Since the Noether current of a global symmetry is essentially determined by its localized version, as we saw in the previous section, we expect that relations between local symmetries will be reflected in relations between the corresponding Noether currents.

To see how this comes about, consider two sets of symmetries of the same action, characterized by infinitesimal parameters $\eps^\alpha_1$ and $\eps^a_2$.\footnote{The below-outlined argument follows closely the presentation in Ref.~\onlinecite{Brauner:2014aha}.} Suppose that the two classes of symmetries are locally \emph{identical}, that is, there is a set of coefficients $f^a_\alpha(x)$ such that a transformation of the first type is equivalent to a transformation of the second type with the choice $\eps^a_2(x)=f_\alpha^a(x)\eps^\alpha_1(x)$. The variation of the action under $\eps_2^a(x)$ reads, by Eq.~\eqref{master},
\begin{equation}
\delta S=\int\dd x\,J^\mu_{2a}\de_\mu\eps_2^a=\int\dd x\,J^\mu_{2a}(f_\alpha^a\de_\mu\eps^\alpha_1+\eps^\alpha_1\de_\mu f_\alpha^a),
\end{equation}
which, due to the assumed local equivalence of the two symmetries, should be equal to
\begin{equation}
\delta S=\int\dd x\,J^\mu_{1\alpha}\de_\mu\eps_1^\alpha.
\end{equation}
This is only possible if
\begin{equation}
J^\mu_{2a}\de_\mu f_\alpha^a=\de_\mu N^\mu_\alpha
\label{consistency}
\end{equation}
for some vector function $N^\mu_\alpha(x)$; we will refer to Eq.~\eqref{consistency} as the ``integrability condition.'' Integration by parts then leads to a linear relation between the currents,
\begin{equation}
J^\mu_{1\alpha}=f_\alpha^aJ^\mu_{2a}-N^\mu_\alpha,
\label{constraint}
\end{equation}
modulo the ambiguity due to adding a vector function whose divergence vanishes off-shell.

Equation~\eqref{constraint} is our main result. We stress, however, that Eq.~\eqref{consistency} imposes a nontrivial constraint, as it has to hold off-shell. As such, it is not a mere consequence of Eq.~\eqref{constraint} and current conservation. In fact, by taking a divergence of Eq.~\eqref{constraint} and using the integrability condition~\eqref{consistency}, we obtain another off-shell identity,
\begin{equation}
\de_\mu J^\mu_{1\alpha}=f^a_\alpha\de_\mu J^\mu_{2a}.
\end{equation}
This underlines the close relation between the conservation laws stemming from the two symmetries.

The correction term~$N_\alpha^\mu$ spoils somewhat the elegance of the result~\eqref{constraint}, which otherwise copies the relation between the associated symmetry transformations. It is therefore natural to ask whether the ambiguity in the definition of the currents could be exploited to remove~$N_\alpha^\mu$. This is unfortunately not always the case, as shown by an explicit counterexample in Sec.~\ref{sec:galileon}. 


\section{Gauge invariance approach}
\label{sec:gauge}

The Noether current can also be derived following an alternative approach that likewise makes use of the local symmetry transformation. The starting point is the \emph{assumption} that the Lagrangian density can be modified by adding a vector field $A_\mu(x)$ so that the action is invariant under the local rather than just global transformations. Technically, this amounts to assuming that the original action $S[\phi]$ is replaced with a new action $\tilde S[\phi,A]$ such that $\tilde S[\phi,0]=S[\phi]$ and that the new action is invariant under a simultaneous transformation of the original, ``matter'' fields $\phi_A$ as in Eq.~\eqref{transfo2} as well as of the gauge field $A_\mu$. We further assume that the transformation of the gauge field takes the generic form
\begin{equation}
A'_\mu(x)=A_\mu(x)+\eps(x)\Xi_\mu(x)+\Sigma(x)\de_\mu\eps(x);
\label{Atransfo}
\end{equation}
the notation is chosen to resemble that used in Eq.~\eqref{transfosigma}.

The variation of the new action to first order in $\eps(x)$ consists of contributions from varying $\phi_A(x)$ and $A_\mu(x)$,
\begin{align}
\delta\tilde S&=\delta_\phi\tilde S+\delta_A\tilde S\\
\notag
&=\delta_\phi\tilde S+\int\dd x\,\frac{\delta\tilde S}{\delta A_\mu(x)}\bigl[\eps(x)\Xi_\mu(x)+\Sigma(x)\de_\mu\eps(x)\bigr].
\end{align}
Upon setting $A_\mu=0$, the term $\delta_\phi\tilde S$ must reproduce the variation of the action in the ungauged theory, Eq.~\eqref{master}. The assumed gauge invariance of $\tilde S$ then implies that there must be a vector function $R^\mu(x)$ such that
\begin{equation}
\frac{\delta\tilde S}{\delta A_\mu(x)}\Xi_\mu(x)\biggr|_{A=0}=-\de_\mu R^\mu(x).
\label{gaugeK}
\end{equation}
An explicit expression for the Noether current follows in turn,
\begin{equation}
J^\mu(x)=-\Sigma(x)\frac{\delta\tilde S}{\delta A_\mu(x)}\biggr|_{A=0}-R^\mu(x).
\label{master_gauged}
\end{equation}

The alternative approach presented here is completely equivalent to the derivation of the Noether current given in Sec.~\ref{sec:intro}, as long as the global symmetry in question can be gauged. As a concrete application of this formalism, we will rederive the main result of Sec.~\ref{sec:relations}, Eq.~\eqref{constraint}. To that end, introduce a set of gauge fields $A_\mu^I(x)$ and parameterize their transformation under the two classes of symmetries with infinitesimal parameters $\eps_1^\alpha(x)$ and $\eps_2^a(x)$ by the functions $\Xi^{I1}_{\alpha\mu}(x)$, $\Sigma^{I1}_\alpha(x)$ and $\Xi^{I2}_{a\mu}(x)$, $\Sigma^{I2}_a(x)$. Assuming that the gauge transformations of $A_\mu^I(x)$ are identical provided one identifies the infinitesimal parameters as $\eps_2^a(x)=f^a_\alpha(x)\eps^\alpha_1(x)$,\footnote{This is a nontrivial assumption, requiring more than just that the two physical global symmetries are locally identical. It demands implicitly that the two symmetries can be simultaneously gauged by adding a single set of gauge fields.} we obtain from Eq.~\eqref{Atransfo}
\begin{equation}
\begin{split}
A'^I_\mu-A^I_\mu&=\eps_2^a\Xi^{I2}_{a\mu}+\Sigma^{I2}_a\de_\mu\eps_2^a\\
&=\eps^\alpha_1\bigl(f^a_\alpha\Xi^{I2}_{a\mu}+\Sigma^{I2}_a\de_\mu f^a_\alpha\bigr)+\Sigma^{I2}_af^a_\alpha\de_\mu\eps_1^\alpha.
\end{split}
\end{equation}
We thus identify the coefficient functions as $\Sigma^{I1}_\alpha=f^a_\alpha\Sigma^{I2}_a$ and $\Xi^{I1}_{\alpha\mu}=f^a_\alpha\Xi^{I2}_{a\mu}+\Sigma^{I2}_a\de_\mu f^a_\alpha$. The condition~\eqref{gaugeK} applied to the first symmetry transformation then takes the form
\begin{equation}
\frac{\delta\tilde S}{\delta A^I_\mu}f^a_\alpha\Xi^{I2}_{a\mu}\biggr|_{A=0}+\frac{\delta\tilde S}{\delta A^I_\mu}\Sigma^{I2}_a\de_\mu f^a_\alpha\biggr|_{A=0}=-\de_\mu R^\mu_{1\alpha}.
\end{equation}
Using Eqs.~\eqref{gaugeK} and~\eqref{master_gauged} applied to the second symmetry transformation, this can be rewritten as
\begin{equation}
\begin{split}
-\de_\mu R^\mu_{1\alpha}&=-f^a_\alpha\de_\mu R^\mu_{2a}-(J^\mu_{2a}+R^\mu_{2a})\de_\mu f^a_\alpha\\
&=-\de_\mu(f^a_\alpha R^\mu_{2a})-J^\mu_{2a}\de_\mu f^a_\alpha.
\end{split}
\end{equation}
Consistency now leads to the integrability condition~\eqref{consistency}, whence we obtain $R^\mu_{1\alpha}=f^a_\alpha R^\mu_{2a}+N^\mu_\alpha$. Plugging this back into Eq.~\eqref{master_gauged} together with the relation $\Sigma^{I1}_\alpha=f^a_\alpha\Sigma^{I2}_a$ then finally reproduces the relation~\eqref{constraint} among the currents.


\section{Examples}
\label{sec:examples}

In order to illustrate the results and arguments of the preceding sections, we will now work out several concrete examples of internal and spacetime symmetries. For the sake of simplicity, we will mostly consider theories whose Lagrangian density depends just on the fields and their first derivatives.


\subsection{Shift symmetries of a massless scalar} 
\label{sec:galileon}

Consider the theory of a free massless scalar field as given by the Lagrangian
\begin{equation}
\La=\frac12(\de_\mu\phi)^2.
\label{gallag}
\end{equation}
Its action is invariant under the following coordinate-dependent transformation,
\begin{equation}
\phi'(x)=\phi(x)+a+b_\alpha x^\alpha,
\end{equation}
known as the ``Galileon'' symmetry.\cite{Nicolis:2008in} For the constant shift parameterized by $a$, we have $\xi=1$ and the invariance condition~\eqref{invariance} is trivially satisfied with $K^\mu=0$. According to Eq.~\eqref{noether}, the corresponding Noether current is $J^\mu=\de^\mu\phi$. On the other hand, for the linear shift parameterized by $b_\alpha$, we have $\xi^\alpha=x^\alpha$ and the invariance condition~\eqref{invariance} is satisfied with $K^\mu_\alpha=\delta^\mu_\alpha\phi$. The Noether current then reads $J^\mu_\alpha=x_\alpha\de^\mu\phi-\delta^\mu_\alpha\phi$ by Eq.~\eqref{noether}.

Observe that the local forms of the two symmetries in this example are identical, as follows by setting $a(x)=x_\alpha b^\alpha(x)$. The integrability condition~\eqref{consistency} where we take $f_\alpha(x)=x_\alpha$ then implies that $N^\mu_\alpha=\delta^\mu_\alpha\phi$. The current $J^\mu_\alpha=x_\alpha\de^\mu\phi-\delta^\mu_\alpha\phi$ then follows from Eq.~\eqref{constraint} at once.

This example also shows that it is not always possible to remove the correction term $N^\mu_\alpha$ in Eq.~\eqref{constraint} by exploiting the ambiguity in the definition of the Noether currents. Indeed, suppose that there \emph{were} improved currents $\tilde J^\mu(x)$ and $\tilde J^\mu_\alpha(x)$ such that $\tilde J^\mu_\alpha(x)=x_\alpha\tilde J^\mu(x)$ holds. By taking the divergence and using the assumed on-shell conservation of both currents, we arrive at the condition,
\begin{equation}
0=\de_\mu\tilde J^\mu_\alpha=\tilde J_\alpha+x_\alpha\de_\mu\tilde J^\mu=\tilde J_\alpha,
\end{equation}
which makes both currents identically vanish on-shell.

Finally, let us use this example to illustrate the gauge invariance method presented in Sec.~\ref{sec:gauge}. The gauged version of the Lagrangian density~\eqref{gallag} reads
\begin{equation}
\La=\frac12(\de_\mu\phi-A_\mu)^2,
\end{equation}
and is invariant under a simultaneous local transformation of the scalar and the gauge field, 
\begin{equation}
\phi'(x)=\phi(x)+\eps(x),\qquad
A'_\mu(x)=A_\mu(x)+\de_\mu\eps(x).
\label{gauge_transfo}
\end{equation}
The transformation of the gauge field has the form~\eqref{Atransfo} with $\Xi_\mu=0$ and $\Sigma=1$. The condition~\eqref{gaugeK} is then trivially satisfied with $R^\mu=0$, and the general result for the current~\eqref{master_gauged} yields $J^\mu=\de^\mu\phi$.

How can we get the other current, $J^\mu_\alpha$, from the same Lagrangian and gauge transformation? We must match Eq.~\eqref{gauge_transfo} to a localized version of the linear shift with parameter $b_\alpha$, i.e.~to identify $\eps(x)=x_\alpha b^\alpha(x)$. The transformation of the gauge field then becomes
\begin{equation}
A'_\mu(x)=A_\mu(x)+b_\mu(x)+x_\alpha\de_\mu b^\alpha(x).
\end{equation}
Matching this to Eq.~\eqref{Atransfo} gives $\Xi_\mu^\alpha=\delta^\alpha_\mu$ and $\Sigma_\alpha=x_\alpha$. The consistency condition~\eqref{gaugeK} then gives $R^\mu_\alpha=\delta^\mu_\alpha\phi$, upon which Eq.~\eqref{master_gauged} reproduces the correct expression for the current, $J^\mu_\alpha=x_\alpha\de^\mu\phi-\delta^\mu_\alpha\phi$.


\subsection{Spacetime translations}
\label{sec:translations}

A shift of the spacetime coordinate, $x'^\mu=x^\mu+a^\mu$, can be written in the form~\eqref{transfo} with $\xi_A=0$ and $\omega^\mu_\alpha=\delta^\mu_\alpha$; the index $\alpha$ labels translations in different directions and $a^\alpha$ plays the role of the parameter of the transformation. The invariance condition~\eqref{invariance} is satisfied with $K^\mu=0$ provided that the Lagrangian density does not depend explicitly on the coordinate, which will be assumed in the following without further mentioning. The corresponding Noether current is the canonical EM tensor,
\begin{equation}
T^{\mu}_{\phantom\mu\alpha}= \delta^\mu_\alpha\La-\PD\La{(\de_\mu\phi_A)}\de_\alpha\phi_A.
\label{MTcanonical}
\end{equation}
A generalization of this result to Lagrangians with derivatives of higher order is trivial. Indeed, upon setting $\xi_{A\alpha}=-\de_\alpha\phi_A$, cf.~Eq.~\eqref{transfo2}, the generalized invariance condition~\eqref{invariance_general} is satisfied with $K^\mu_\alpha=-\delta^\mu_\alpha\La$, and Eq.~\eqref{noether_general} then gives a general expression for the EM tensor,
\begin{align}
&T^{\mu}_{\phantom\mu\alpha}=\delta^\mu_\alpha\La-\sum_{n=1}^\infty\sum_{k=1}^n(-1)^{k+1}\\
\notag
&\times\left[\de_{\mu_2}\dotsb\de_{\mu_k}\frac{\de\La}{\de(\de_{\mu}\de_{\mu_2}\dotsb\de_{\mu_n}\phi_A)}\right](\de_{\mu_{k+1}}\dotsb\de_{\mu_n}\de_\alpha\phi_A).
\end{align}
This important example of a Noether current will be discussed in much more depth in Sec.~\ref{sec:EMtensor}.


\subsection{Spacetime rotations}
\label{sec:rotations}

For the sake of illustration, we now limit our discussion to the simplest case of scalar fields. We can then again set $\xi_A=0$ and take the coordinate transformation as $x'^\mu=x^\mu-\theta^{\mu\nu}x_\nu=x^\mu+\frac12\theta^{\alpha\beta}\omega^\mu_{\alpha\beta}$, where $\theta^{\alpha\beta}$ is an antisymmetric rank-two tensor of infinitesimal transformation parameters and $\omega^\mu_{\alpha\beta}=x_\alpha\delta^\mu_\beta-x_\beta\delta^\mu_\alpha$. The invariance condition~\eqref{invariance} is satisfied with $K^\mu=0$ provided $[\de\La/\de(\de_\mu\phi_A)]\de^\nu\phi_A$, or equivalently $T^{\mu\nu}$, is symmetric in the indices $\mu,\nu$, which is the case for scalar fields.\footnote{Here Lorentz indices are assumed to be lowered by the Minkowski metric $\eta_{\mu\nu}$ and raised by its inverse $\eta^{\mu\nu}$.} Eq.~\eqref{noether} then implies for the conserved tensor current,
\begin{equation}
\begin{split}
M^{\mu\alpha\beta}&=\left[\eta^{\mu\nu}\La-\PD\La{(\de_\mu\phi_A)}\de^\nu\phi_A\right]\left(x^\alpha\delta^\beta_\nu-x^\beta\delta^\alpha_\nu\right)\\
&=x^\alpha T^{\mu\beta}-x^\beta T^{\mu\alpha}.
\end{split}
\end{equation}
This relation between the EM and angular momentum tensors is not accidental. Local infinitesimal translations and rotations coincide provided one identifies their parameters by $a^\mu(x)=\frac12f^\mu_{\alpha\beta}(x)\theta^{\alpha\beta}(x)$, where $f^\mu_{\alpha\beta}=\omega^\mu_{\alpha\beta}=x_\alpha\delta^\mu_\beta-x_\beta\delta^\mu_\alpha$. Next, we note that $T^\mu_{\phantom\mu\nu}\de_\mu f^\nu_{\alpha\beta}=T_{\alpha\beta}-T_{\beta\alpha}$ vanishes provided $T^{\mu\nu}$ is symmetric. The integrability condition~\eqref{consistency} is then satisfied with $N^\mu_{\alpha\beta}=0$, and Eq.~\eqref{constraint} immediately tells us that
\begin{equation}
M^{\mu}_{\phantom\mu\alpha\beta}= f^\nu_{\alpha\beta}T^\mu_{\phantom\mu\nu}=x_\alpha T^\mu_{\phantom\mu\beta}-x_\beta T^\mu_{\phantom\mu\alpha},
\end{equation}
in agreement with the result obtained above directly from Noether's (first) theorem.


\subsection{Dilatations}

For a further illustration, let us return to the theory of a free massless scalar field, Eq.~\eqref{gallag}. An infinitesimal dilatation is a rescaling of the coordinate with infinitesimal parameter $\delta$, $x'^\mu=(1+\delta)x^\mu$, so that $\omega^\mu(x)=x^\mu$. Suppose that the scalar field rescales simultaneously according to
\begin{equation}
\phi'(x')=(1+\delta)^\Delta\phi(x),
\end{equation}
so that $\xi(x)=\Delta\phi$. The invariance condition~\eqref{invariance} is satisfied with $K^\mu=0$ provided that $\Delta=1-d/2$, where $d$ is the dimension of the spacetime. Equation~\eqref{noether} then gives the canonical dilatation current\cite{Treiman:1972}
\begin{equation}
D^\mu=x^\nu T^{\mu}_{\phantom\mu\nu}+\Delta\phi\de^\mu\phi,
\end{equation}
where the canonical EM tensor is given by the right-hand side of Eq.~\eqref{MTcanonical}.

In $d=2$ dimensions, the scale dimension $\Delta$ of the scalar field vanishes and a local dilatation can be recovered from a local translation with $a^\mu(x)=f^\mu(x)\delta(x)$ where $f^\mu(x)=x^\mu$. Moreover, $T^\mu_{\phantom\mu\nu}\de_\mu f^\nu=0$, hence the integrability condition~\eqref{consistency} is trivially satisfied. Equation~\eqref{constraint} then implies that the dilatation current is related to the EM tensor by $D^\mu=f^\nu T^\mu_{\phantom\mu\nu}=x_\nu T^{\mu\nu}$, which agrees with the above explicit calculation. Note that the EM tensor is sometimes ``improved'' \emph{ad hoc} so that the relation $D^\mu=x_\nu T^{\mu\nu}$ holds;~\cite{Treiman:1972} our discussion of locally equivalent symmetries suggests that this relation is only natural in two spacetime dimensions.


\subsection{Galilei transformations}

Until now, most of the discussed examples were explicitly relativistic, and we even used consistently the four-vector notation. However, our general results of course apply equally well to nonrelativistic systems. For a concrete example, consider a theory of a complex Schr\"odinger field $\psi(x)$. This is usually equipped with the symmetry under a change of phase of the field, $\psi'(x)=e^{\imag\theta}\psi(x)$, which implies conservation of the number of particles.

What really distinguishes nonrelativistic systems from relativistic ones, however, is the invariance under Galilei boosts as opposed to Lorentz transformations. Under a coordinate boost $\vek x'=\vek x+\vek vt$ with velocity $\vek v$, the Schr\"odinger field transforms as
\begin{equation}
\psi'(\vek x',t)=e^{\imag m(\vek v\cdot\vek x+\frac12\vek v^2t)}\psi(\vek x,t).
\end{equation}
An infinitesimal boost takes accordingly the form $\psi'(\vek x',t)=\psi(\vek x,t)-\imag mv^ix_i\psi(\vek x,t)$.\footnote{Note that we use the relativistic convention where raising or lowering a spatial index adds a minus sign.} We can thereby identify $\xi_i(x)=-\imag mx_i\psi(x)$ and $\omega^\mu_i(x)=\delta^\mu_it$, where the velocity $v^i$ serves as the parameter of the transformation.

We will not attempt to write down a fully general expression for the Noether current arising from invariance under Galilei boosts. Instead, we will concentrate on its interplay with Noether currents of other symmetries. A local infinitesimal Galilei boost is equivalent to a combination of a local translation with $a^\mu(x)=\delta^\mu_iv^i(x)t$ and a local phase transformation with $\theta(x)=-mx_iv^i(x)$. In the language of Sec.~\ref{sec:relations}, this corresponds to $f^\mu_i(x)=\delta^\mu_it$ and $\tilde f_i(x)=-mx_i$. By Eq.~\eqref{consistency}, the EM tensor $T^{\mu\nu}$ and the particle number current $J^\mu$ must satisfy the integrability condition\cite{Greiter:1989qb}
\begin{equation}
T^{0i}-mJ^i=\de_\mu N^{\mu i}
\label{wingate}
\end{equation}
with some vector function $N^{\mu i}$. By Eq.~\eqref{constraint}, the boost Noether current $B^{\mu i}$ then satisfies the relation
\begin{equation}
B^{\mu i}=tT^{\mu i}-mx^iJ^\mu-N^{\mu i}.
\label{wingate2}
\end{equation}

Whether or not the vector function $N^{\mu i}(x)$ vanishes is a dynamical question that can be answered by an analysis within a concrete model. It can nevertheless be shown that $N^{\mu i}=0$ provided the spatial translation and phase symmetries of the Schr\"odinger field can be simultaneously gauged as required by the approach of Sec.~\ref{sec:gauge}. Namely, by a slight generalization of the argument therein, one can then deduce identities equivalent to Eqs.~\eqref{wingate} and~\eqref{wingate2} with $N^{\mu i}=0$.\cite{Brauner:2014aha} A detailed discussion would, however, take us far beyond the scope of this paper, as it would require the knowledge of the nonrelativistic version of general coordinate invariance.\cite{Son:2005rv}


\section{Energy--momentum tensor}
\label{sec:EMtensor}

The EM tensor is undoubtedly the most important example of a Noether current. In Sec.~\ref{sec:translations}, we derived the canonical EM tensor $T^{\mu\nu}(x)$ using Eq.~\eqref{noether_general}. However, a different definition of the EM tensor is often used when the given set of matter fields is (or can be) coupled to a background metric tensor $g_{\mu\nu}(x)$ in such a way that the resulting action $\tilde S[\phi,g]$ is invariant under general coordinate transformations. The ``metric'' EM tensor is thus defined by
\begin{equation}
\theta^{\mu\nu}(x)=2\frac{\delta\tilde S}{\delta g_{\mu\nu}(x)}\biggr|_{g=\eta},
\label{MTmetric}
\end{equation}
where $\eta_{\mu\nu}$ is the actual (Minkowski) metric of the spacetime. It is not \emph{a priori} obvious how the tensors $T^{\mu\nu}(x)$ and $\theta^{\mu\nu}(x)$ are related, and it is in fact well known that they in general do not coincide, notably when the theory contains fields with spin (see Ref.~\onlinecite{Blaschke:2016ohs} for a review of the various constructions of the EM tensor).

As a simple example, consider the theory of a massive vector field $B_\mu(x)$, defined by
\begin{equation}
\La=-\frac14F_{\mu\nu}F^{\mu\nu}-\frac12m^2B_\mu B^\mu,
\end{equation}
where $F_{\mu\nu}=\de_\mu B_\nu-\de_\nu B_\mu$.\footnote{The special case of $m=0$ is Maxwell's electrodynamics in the absence of sources of electric charge. We choose to work with the general massive case in order to avoid confusion related to gauge invariance.} The field $B_\mu$ is to be identified with the matter field $\phi_A$ in Eq.~\eqref{transfo}, and using the identity $\de\La/\de(\de_\mu B_\nu)=-F^{\mu\nu}$, we obtain from Eq.~\eqref{MTcanonical} the canonical EM tensor,
\begin{equation}
T^{\mu\nu}=\eta^{\mu\nu}\La+F^{\mu\alpha}\de^\nu B_\alpha.
\label{elmagT}
\end{equation}
The metric EM tensor is found by writing the action in a generally covariant form. Skipping the well-known steps of the calculation, we just quote the result,
\begin{equation}
\theta^{\mu\nu}=\eta^{\mu\nu}\La+F^{\mu\alpha}F^\nu_{\phantom\nu\alpha}+m^2B^\mu B^\nu.
\label{elmagtheta}
\end{equation}
Thanks to the antisymmetry of $F^{\mu\nu}$, we have
\begin{equation}
\begin{split}
T^{\mu\nu}&\simeq\eta^{\mu\nu}\La+F^{\mu\alpha}\de^\nu B_\alpha-\de_\alpha(F^{\mu\alpha}B^\nu)\\
&=\theta^{\mu\nu}-B^\nu(\de_\alpha F^{\mu\alpha}+m^2B^\mu),
\end{split}
\end{equation}
where the symbol $\simeq$ indicates equality up to a term whose divergence identically vanishes off-shell. The expression in the parentheses on the second line defines the equation of motion for the vector field $B^\mu$, and thus vanishes on-shell. The $T^{\mu\nu}$ and $\theta^{\mu\nu}$ tensors are therefore related, but not equal or even physically equivalent.

The metric EM tensor~\eqref{elmagtheta} is distinguished from the canonical EM tensor~\eqref{elmagT} by two appealing properties: the symmetry in $\mu,\nu$ and the dependence solely on $F^{\mu\nu}$ in the limit $m\to0$. The problem how to ``improve'' the canonical EM tensor in order to reproduce these properties has a long history, from a purely \emph{ad hoc} argument\cite{Belinfante:1940} to more refined approaches, typically invoking additional constraints to fix the ambiguity in the EM tensor, especially the electromagnetic gauge invariance.\cite{Jackiw:1978ar,*Eriksen:1979vq,*Zhu:1980dy,*Munoz:1996wp}

Here we wish to understand the relation between the canonical and metric EM tensors using only translational invariance, building on our understanding of the intimate connection between the form of the Noether current and the local symmetry transformation, used to obtain it via Eq.~\eqref{master}. Naively, the metric EM tensor is exactly what we would expect from the gauge-invariance-based approach of Sec.~\ref{sec:gauge}, which we claimed therein to be merely an alternative way to derive the same canonical Noether current as in Sec.~\ref{sec:intro}. How is it then \emph{possible} that the canonical and metric EM tensors do not coincide?

This naive expectation has two flaws, which suggest two possibilities how to reconcile the canonical and metric EM tensors. First, under general coordinate transformations, a vector field (in fact, any non-scalar field) does not transform according to Eq.~\eqref{transfo} with $\xi_A=0$, which was the starting point in the derivation of the canonical EM tensor in Sec.~\ref{sec:translations}. Second, the transformation of the metric $g_{\mu\nu}$ under general coordinate transformations does not match the ansatz~\eqref{Atransfo}, on which our gauge invariance approach in Sec.~\ref{sec:gauge} is based. Below we will show that: (i) it is perfectly possible to obtain the symmetric EM tensor $\theta^{\mu\nu}$ using Noether's canonical approach, provided that we start from a suitably defined local translation; (ii) it is perfectly possible to reproduce the canonical EM tensor $T^{\mu\nu}$ using the gauge invariance approach, provided that we express the generally covariant action solely in terms of scalar field variables, whose transformation properties match Eq.~\eqref{transfo} with $\xi_A=0$.


\subsection{Metric EM tensor from Noether's theorem}

Under the infinitesimal general coordinate transformation $x'^\mu=x^\mu+a^\mu(x)$, a covariant vector field such as $B_\mu(x)$ transforms as
\begin{equation}
B'_\mu(x')=B_\mu(x)-B_\nu(x)\de_\mu a^\nu(x).
\label{star}
\end{equation}
This does not match Eq.~\eqref{transfo}, which does not contain any derivatives of the transformation parameter. We therefore have to resort to its generalization, Eq.~\eqref{transfosigma}. Upon rewriting Eq.~\eqref{star} as $B'_\mu(x')=B_\mu(x)+\sigma^\nu_{\mu\alpha}(x)\de_\nu a^\alpha(x)$, we can identify $\sigma_{\mu\alpha}^\nu(x)=-\delta^\nu_\mu B_\alpha(x)$. This gives via Eq.~\eqref{Jsigma} the canonical EM tensor, corresponding to the local translation of the field, defined by Eq.~\eqref{star},
\begin{equation}
T^{\mu\nu}= T^{\mu\nu}\Bigr|_{\sigma=0}+\eta^{\nu\alpha}\sigma^\mu_{\lambda\alpha}\frac{\delta S}{\delta B_\lambda}\simeq\theta^{\mu\nu},
\end{equation}
where we used the fact that $\delta S/\delta B_\mu=-(\de_\alpha F^{\mu\alpha}+m^2B^\mu)$ in the last step.

This shows that the metric EM tensor $\theta^{\mu\nu}$ can be naturally recovered from the canonical Noether procedure. There is no ambiguity in the derivation apart from that expressed by the $\simeq$ relation; we merely insisted that the local form of the symmetry transformation, used to deduce the EM tensor via Eq.~\eqref{master}, matches the form of the general coordinate transformation of the field.


\subsection{Canonical EM tensor from gauged action}

We would now like to apply the formalism developed in Sec.~\ref{sec:gauge} to an action where spacetime translations are gauged by coupling the matter fields to a background spacetime geometry. To that end, we trade the metric $g_{\mu\nu}(x)$ for the vielbein $e^\mu_a(x)$. This constitutes a set of fixed vectors, labeled by the index $a$, forming a local basis at any spacetime point. It is customary, though not mandatory, to require that the basis be orthonormal, that is $g_{\mu\nu}e^\mu_ae^\nu_b=\eta_{ab}$. The vector index $\mu$ of the vielbein is raised and lowered using the spacetime metric $g_{\mu\nu}$, whereas the basis index $a$ is raised and lowered using the Minkowski metric $\eta_{ab}$. The inverse relation $g_{\mu\nu}=\eta_{ab}e^a_\mu e^b_\nu$, which expresses the completeness of the vielbein basis, allows us to trade the metric for the covariant vielbein vectors $e^a_\mu(x)$.

Under the infinitesimal general coordinate transformation $x'^\mu=x^\mu+a^\mu(x)$, the covariant vielbein transforms just like the $B_\mu$ field in Eq.~\eqref{star}. This can be expressed equivalently as 
\begin{equation}
e'^a_\mu(x)=e^a_\mu(x)-e^a_\alpha\de_\mu a^\alpha-a^\alpha\de_\alpha e^a_\mu,
\end{equation}
which takes the form of Eq.~\eqref{Atransfo}. We thus identify
\begin{equation}
\Xi^a_{\alpha\mu}=-\de_\alpha e^a_\mu,\qquad
\Sigma^a_\alpha=-e^a_\alpha.
\end{equation}
The absence of the background gauge field corresponds to $e^a_\mu(x)=\delta^a_\mu$ so that the consistency condition~\eqref{gaugeK} is satisfied trivially with $R^\mu_\alpha=0$. The energy-momentum tensor is then given by Eq.~\eqref{master_gauged} as
\begin{equation}
T^\mu_{\phantom\mu\alpha}(x)=e^a_\alpha(x)\frac{\delta\tilde S}{\delta e^a_\mu(x)}\biggr|_{e=\delta}.
\label{gaugeeta}
\end{equation}
As follows from the derivation in Sec.~\ref{sec:gauge}, this EM tensor has to equal that extracted from the canonical procedure based on Eq.~\eqref{master}, provided that the local symmetry transformations of the matter fields $\phi_A$ in the two approaches match. To that end, it is important to realize that different local symmetry transformations of the matter fields require different forms of the generally coordinate invariant action $\tilde S[\phi,e]$.

The metric EM tensor $\theta^{\mu\nu}$ is obtained via Eq.~\eqref{gaugeeta} from a generally covariant action where the matter field $B_\mu$ is a covariant vector, that is transforms according to Eq.~\eqref{star}. Such an action can be taken as
\begin{align}
\label{elmagaction1}
\tilde S[B_\mu,e^a_\mu]={}&\int\dd x\sqrt{-\|g\|}\\
\notag
&\times\bigl(-\tfrac14g^{\mu\alpha}g^{\nu\beta}F_{\mu\nu}F_{\alpha\beta}-\tfrac12m^2g^{\mu\nu}B_\mu B_\nu\bigr),
\end{align}
where $g^{\mu\nu}=\eta^{ab}e^\mu_ae^\nu_b$. In order to compute the variation of the action with respect to the covariant vielbein $e^a_\mu$, one can use the identities $\delta e^\mu_a=-e^\mu_be^\nu_a\delta e^b_\nu$ and $\delta\|e\|=\|e\|e^\mu_a\delta e^a_\mu$. Eq.~\eqref{gaugeeta} then yields the EM tensor~\eqref{elmagtheta} after a straightforward computation.

The nonsymmetric EM tensor $T^{\mu\nu}$~\eqref{elmagT} can likewise be obtained via Eq.~\eqref{gaugeeta} from a generally covariant action where the matter field is a \emph{scalar}, that is, transforms according to  Eq.~\eqref{transfo} with $\xi=0$. This can be achieved by projecting $B_\mu$ on the vielbein to get $B_a(x)=e^\mu_a(x)B_\mu(x)$, and analogously $F_{ab}=e^\mu_a\de_\mu B_b-e^\mu_b\de_\mu B_a$. In this case, the generally covariant action can thus be taken as
\begin{equation}
\tilde S[B_a,e^a_\mu]=\int\dd x\,\|e\|\bigl(-\tfrac14\eta^{ac}\eta^{bd}F_{ab}F_{cd}-\tfrac12m^2\eta^{ab}B_aB_b\bigr).
\end{equation}
It is then a matter of a straightforward exercise to show that Eq.~\eqref{gaugeeta} indeed leads to the canonical EM tensor~\eqref{elmagT}.


\section{Summary and conclusions}

The (first) Noether theorem predicts the existence of a conserved current in a system with a global continuous symmetry. In this article, we showed that additional insight may be gained by inspecting the role of the corresponding local field transformations. While not being a symmetry of the action on their own, these provide a useful tool to identify the Noether current. Moreover, they illuminate part of the ambiguity associated with the definition of the current.

The emphasis on local symmetry transformations led us naturally to certain linear relations between currents of distinct global symmetries that coincide on the local level. While none of the examples worked out in Sec.~\ref{sec:examples} is new, our general argument provides a unified framework for their understanding. Finally, our analysis of the correspondence between different local symmetry transformations associated with the same global symmetry, and different forms of the Noether current of this symmetry, sheds (what we believe to be) new light on the infamous discrepancy between the canonical and metric EM tensors in theories of particles with spin.


\begin{acknowledgments}
We acknowledge financial support from the ToppForsk-UiS program of the University of Stavanger and the University Fund, Grant No.~PR-10614.
\end{acknowledgments}


\bibliography{references}

\end{document}